\documentclass[a4paper]{jpconf}
\usepackage{graphicx}

\def\dis{distribution}
\def\pt{p_T}
\def\bq{\begin{eqnarray}}
\def\eq{\end{eqnarray}}
\def\np{N_{\rm part}}
\def\ph{$\phi$}
\def\om{$\Omega$}
\def\ptt{$p_T$}

\begin{document}

\title{Phenomenological Implications of the $p_T$ Spectra of  $\phi$ and $\Omega$ produced at LHC and RHIC}

\author{Rudolph C. Hwa$^1$ and Lilin Zhu$^2$}

\address{$^1$Institute of Theoretical Science and Department of Physics University of Oregon, Eugene, OR 97403-5203, USA}
\address{$^2$ Department of  Physics, Sichuan University, Chengdu  610064, P.\ R.\ China}

\ead{hwa@uoregon.edu}

\begin{abstract}
The data on the $p_T$ spectra of $\phi$ and $\Omega$ at LHC can be presented in a format that shows exponential behavior up to $p_T\approx 6$ GeV/c with the same slope for both particles and for nearly all centralities. They are empirical properties that are shared at lower energies with the inverse slope showing a power-law dependence on $\sqrt{s_{NN}}$. The shared properties of the spectra are shown to emerge naturally from the recombination model. No flow is needed. We find experimental hints for the possibility  that  $\phi$ and $\Omega$  are mostly produced in the ridge that are generated by minijets. Appropriate experimental test is suggested.
\end{abstract}

The production of \ph\ and \om\ at LHC with \ptt\ reaching as high as 5 GeV/c (for \ph) \cite{al1} and 7 GeV/c (for \om) \cite{al2} has not been discussed in any theoretical framework ranging from hydrodynamics to QCD. Both the extent of the  \ptt\ range and the constituents being only $s$ quarks present challenges to the usual theories. Because of those difficulties  we find it opportune to investigate the subject in a novel phenomenological way with the hope of discovering clues to what may be flaws in the conventional interpretation of what happens in heavy-ion collisions at very high energies.

The \ptt\ spectra of  \ph\ have been fitted in  \cite{al1} by blast wave in the hydro model \cite{bl} using 3 free parameters for each centrality bin. Undoubtedly, the same can be done for the \om\ spectra. So what does one learn from the 6 parameters for each centrality? We propose a different way to examine the \ptt\ spectra. The usual presentation of data is in terms of $dN_h/\pt d\pt$. Let us define for meson and baryon data, respectively, the two functions
\bq
M_h(\pt)={dN_h\over \pt d\pt}\cdot {m^h_T\over p_T} ,  \label{1} \\
B_h(\pt)={dN_h\over \pt d\pt}\cdot {m^h_T\over p^2_T} ,  \label{2}
\eq
which are just different ways of  presenting of the same data without theoretical adjustments. Note that they have different dimensions. In Figs.\ 1 and 2 we show the \dis s $M_\phi(\pt)$ and $B_\Omega(\pt)$ using only the data points in  \cite{al1,al2}. They show the strikingly simple exponential behaviors in \ptt\ for nearly all centralities. The straight-line fits of both \dis s amazingly have the same inverse slope
\bq
T_\phi = T_\Omega = 0.51\ {\rm GeV/c} , \quad {\rm for}\ \sqrt{s_{NN}}=2.76\ {\rm TeV}  \label {3}
\eq
a property that is totally hidden in the original data.

\begin{figure}[h]
\begin{minipage}{18pc}
\includegraphics[width=3.3in]{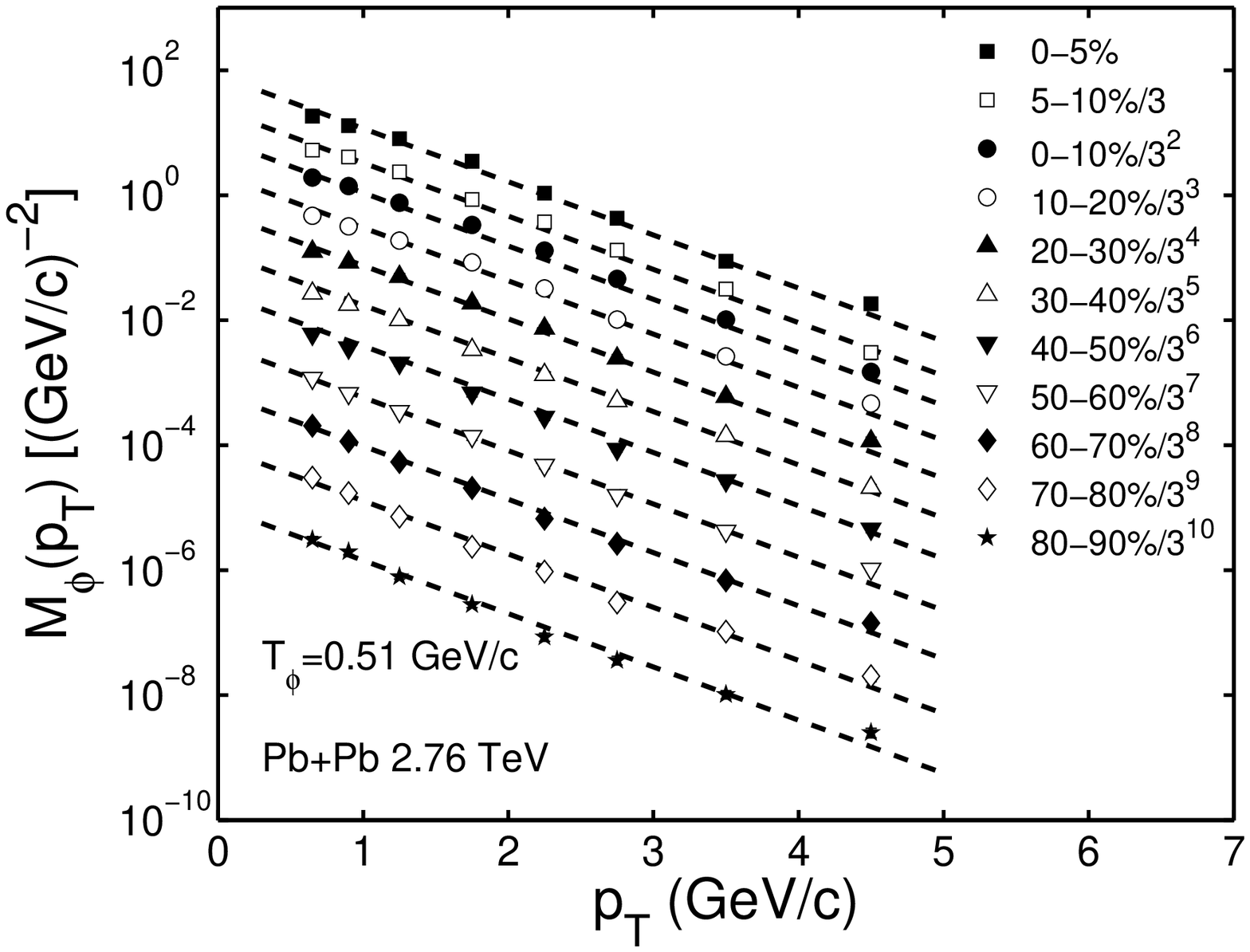}
\caption{\label{label} Plot of $M_\phi(\pt)$ from data in \cite{al1}.}
\end{minipage}\hspace{2pc}%
\begin{minipage}{18pc}
\includegraphics[width=3.3in]{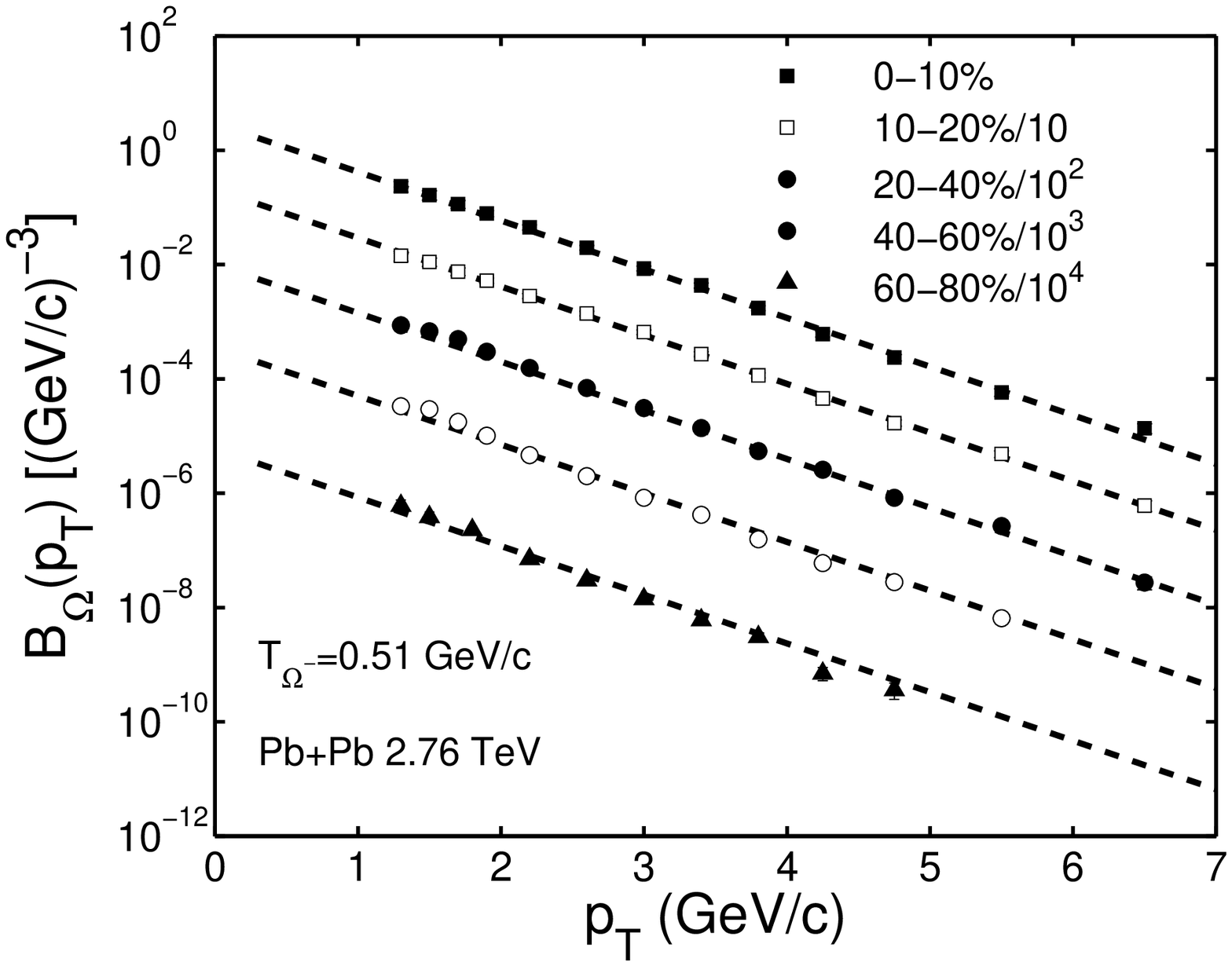}
\caption{\label{label}Plot of $B_\Omega(\pt)$ from data in \cite{al2}.}
\end{minipage} 
\end{figure}

Seeing two different functions of different dimensions having the same inverse slope in semi-log plots stimulates curiosity. Exponential behavior in \ptt\ usually means thermal \dis, but being valid out to $\pt\approx 6$ GeV/c is not what QCD thermodynamics can offer to explain. The fits in Figs.\ 1 and 2 can be expressed in the factorizable form
\bq
M_\phi(\pt, \np)=A_\phi(\np) \exp(-\pt/T_\phi) , \label{4}  \\
B_\Omega(\pt, \np)=A_\Omega(\np) \exp(-\pt/T_\Omega) , \label{5} 
\eq
where the centrality dependences for $\np>70$ can  be shown to satisfy
\bq
A_{\phi,\Omega}(\np)=A^0_{\phi,\Omega} \np^{a_{\phi,\Omega}} ,\quad a_\phi=0.9, \quad a_\Omega=1.35. \label{6}
\eq

What is found above for $\sqrt{s_{NN}}=2.76$ TeV at LHC turns out to be also true at lower energies with different inverse slopes down to 7.7 GeV at RHIC \cite{rhphi, rhomega130, rhomega200, bes}. In Fig.\ 3 we show how $T_{\phi,\Omega}$ depends on  $\sqrt{s_{NN}}$ empirically. The extension of the power-law behavior to 5.02 TeV yields $T_{\phi,\Omega}=0.54$ GeV/c, which is a prediction without theory. We note that the values of $T_{\phi,\Omega}$  shown in Fig.\ 3 are not the temperatures discussed in  hydro models.

Now, we summarize briefly how the observed behaviors of $M_\phi$ and $B_\Omega$ can be understood in the recombination model, a review of which can be found in the first few sections of \cite{reco}. The invariant \dis\ of $\phi$ at mid-rapidity is
\bq
E{dN_\phi\over d\pt}=\int {dp_1\over p_1}{dp_2\over p_2}{\cal T}_s(p_1){\cal T}_{\bar s}(p_2)R_\phi(p_1,p_2,\pt) ,   \label{7}
\eq
where only the thermal-parton \dis s, ${\cal T}(p_i)$, are retained, since the production of strange shower partons is suppressed. The recombination function $R_\phi(p_1,p_2,\pt)$ contains a momentum-conservation factor $\delta((p_1+p_2)/\pt -1)$. Inserting into (7)  ${\cal T}_s(p_1)=p^0_1dN_s/dp_1=C_sp_1e^{-p_1/T_s}$
where $T_s$ is the inverse slope for $s$ quark \dis, and a similar form for
  ${\cal T}_{\bar s}(p_2)$, yields (with the assumption $T_s=T_{\bar s}$)
\bq
E{dN_\phi\over d\pt}=C_sC_{\bar s}Z_\phi \pt^2 e^{-p_T/T_s} , \label{8}
\eq
where $Z_\phi$ is a numerical factor arising from the integration. At $y\approx 0$ where $E\approx m_T$, Eq.\ (8) can be put into the form for $M_\phi$  given in (4)  with $A_\phi(\np)=C_sC_{\bar s}Z_\phi $ and $T_\phi=T_s$.

\begin{figure}[h]
\begin{minipage}{18pc}
\includegraphics[width=3.2in]{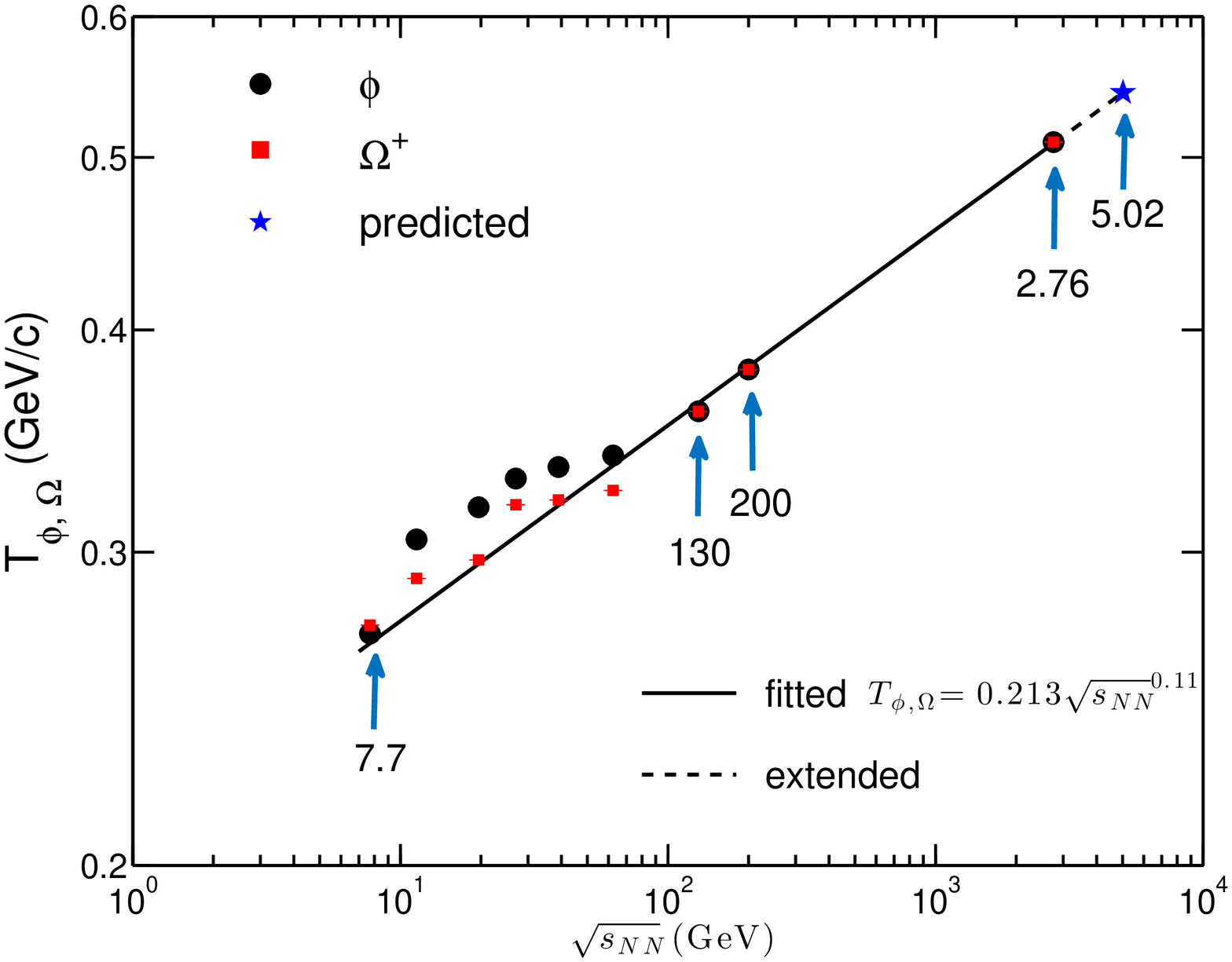}
\caption{\label{label}$T_{\phi,\Omega}$ derived from \cite{al1,al2,rhphi, rhomega130, rhomega200,bes}.}
\end{minipage}\hspace{2pc}%
\begin{minipage}{18pc}
\includegraphics[width=3.2in]{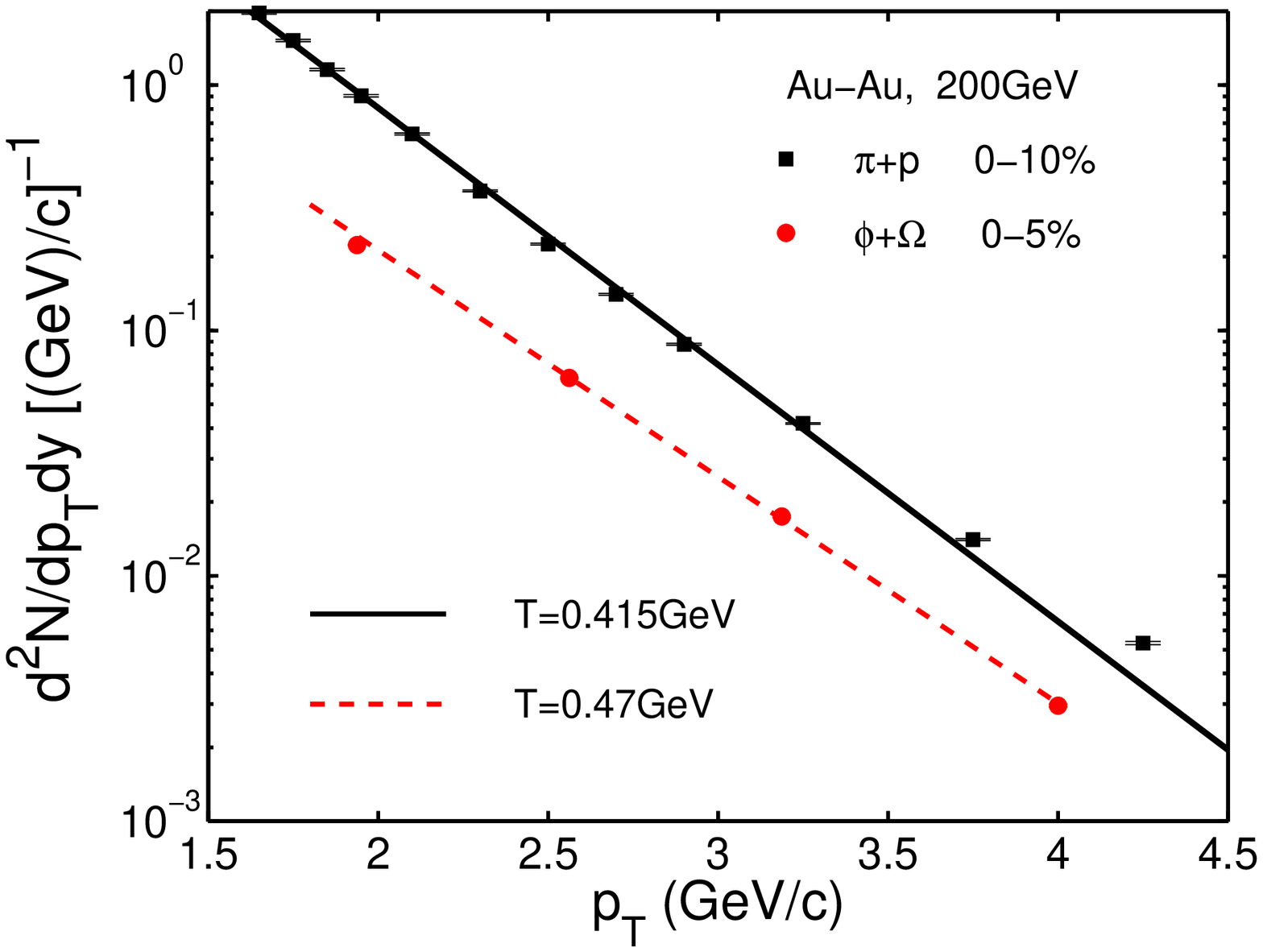}
\caption{\label{label}RHIC data on $\pi+p$ and $\phi+\Omega$ \cite{rhphi, rhomega200, phenix}.}
\end{minipage} 
\end{figure}

For $\Omega$ production three ${\cal T}_{\bar s}(p_i)$ are involved in $E{dN_\Omega/d\pt}$, so one obtains (5) with $A_\Omega(\np)=C_s^3Z_\Omega $ and $T_\Omega=T_s$. Comparing $M_\phi$ with $B_\Omega$, one gets $T_\phi=T_\Omega=T_s$ and that the ratio $A_\phi(\np)/A_\Omega(\np)$ is proportional to $C_sC_{\bar s}/C_s^3$, which agrees with the empirical $\np$ dependence in (6) at LHC, i.e., 2:3 in the exponents, in the reasonable assumption  that $C_{\bar s}$ depends on $\np$ as $C_s$ does.

What is important in our description above is that we have made no mention of flow. There is no need for radial velocity, $T_{\rm kin}$ or blast wave. The system must expand in some way, but it need not be prescribed by hydrodynamics without jets.  We have for nearly all centralities used only one parameter $T_s$, which we have referred to as inverse slope, not temperature, although ${\cal T}_s(p_i)$ is regarded as thermal \dis. Since hadronization by recombination occurs at the end of the evolutionary process, we have not made use of any specific expansion dynamics. ${\cal T}_s(p_i)$ is the invariant $s$ quark \dis\ at the time of hadronization 
without any contribution from the shower partons that arise from fragmentation of high-$\pt$ jets. But at LHC minijets are copiously produced and must influence the bulk of the medium that hydro ignores. In our view ${\cal T}_s(p_i)$ must contain the effects of minijets.

The most notable feature about the above description is that the common value of the inverse slopes in Figs.\ 1 and 2 is $T_{\phi,\Omega}=0.51$ GeV/c, as given in (3) for $\sqrt{s_{NN}}=2.76$ TeV. Even at 7.7 GeV in the BES program the value is $T_{\phi,\Omega}=0.27$ GeV/c, still much larger than the chemical freeze-out temperature of $T_{\rm ch}<0.165$ GeV and kinetic freeze-out temperature of $T_{\rm kin}<0.14$ GeV \cite{bm}. The fact that the exponential fits of $M_{\phi}(\pt)$ and $B_{\Omega}(\pt)$ are so good for such a wide range of \ptt\ demands an explanation. An equation such as (8) relates $T_s$ to the observed value of $T_{\phi,\Omega}$ but does not explain why $T_s$ is so high. Our first step  is to ask how high $T_s$ is relative to the inverse slope for light quarks. It turns out that, when (1) is applied to pions, $M_\pi(\pt)$ cannot be fitted by a simple exponential because the fragmentation products and resonance decay contributions bend the \dis\ above a straight line at both high and low $\pt$ regions. The situation is better for $B_p(\pt)$ for proton, but still not as good as in Fig.\ 2. Nevertheless, the common straight-line fits of $M_\pi(\pt)$ and $B_p(\pt)$ yields $T_{\pi,p}=0.41$ GeV/c. That gives us an estimate of the difference between $T_\Omega$ and $T_p$: $\Delta T=T_\Omega-T_p=0.1$ GeV/c at LHC. A similar comparison at RHIC-200 yields $\Delta T=0.065$ GeV/c.

That leads us to recall the difference in inverse slopes when ridge was first discovered by STAR \cite{jp}: $T_{\rm ridge}-T_{\rm incl}$=40-50 MeV. We use the STAR data for Au-Au collisions at 200 GeV and compare $d^2N/d\pt dy$ at mid-rapidity between $\pi+p$ (0-10\%) and $\phi+\Omega$ (0-5\%) as shown in Fig.\ 4. In the range between $\pt=1.5$ and 4 GeV/c, which is the range on ridge in \cite{jp}, the data can be well fitted by straight lines, showing the difference in inverse slopes to be $T_{\phi+\Omega}-T_{\pi+p}=55 $ MeV. That gives us a strong hint that $\phi$ and $\Omega$ may be a component in the ridge, but instead of being in the triggered events, they are identified.

So far there are no data from LHC on the $\pt^{\rm assoc}$ dependence of the particles in the ridge. However, CMS does have evidence that the ridge yield increases with decreasing $\pt^{\rm trig}$ \cite{cms}. Since a trigger particle comes from the fragmentation of a hard parton that emerges from the expanding medium, lower $\pt^{\rm trig}$  is more likely to be associated with a hard parton originated from the interior of the medium, thus losing a larger portion its energy traversing that medium. The energy lost enhances the thermal partons, so the ridge that comes from the enhanced thermal partons \cite{reco,hz} gets higher inverse slope and thus more yield. To confirm our interpretation that $\phi$ and $\Omega$ are in the ridge with higher $T_{\phi,\Omega}$ than $T_{\pi,p}$, we need data from LHC to identify \ph\ and \om\ in the triggered events and to compare the inverse slopes of their $\pt^{\rm assoc}$ \dis s  with (3).

In the usual QCD thermodynamics the chemical and kinetic freeze-out temperatures are less than 160 MeV. The larger $\left<\pt\right>$, corresponding to our much larger value of inverse slope $T_{\phi,\Omega}$, is then achieved in hydro models by transverse flow, which is a description predicated on the assumption that the system evolving smoothly from early times. But we know from the parton \dis\ functions $F(x)$ that they increase steeply with decreasing momentum fraction $x$, so in an AA collision the probability is high for scattering among low-$x$ partons resulting in a preponderance of minijets which can propagate inside the expanding medium  and lose energy in ways that invalidate the rapid thermalization hypothesis. Minijets with $\pt\sim$ 2-3 GeV/c cannot be calculated reliably in pQCD, but their effects cannot be ignored. They are almost all absorbed by the environment, resulting in enhancement of the thermal partons that creates the ridge phenomenon \cite{hz,hw}. The minijets can produce more $s\bar s$ pairs than in equilibrium QCD thermodynamics. It is therefore natural for us to interpret the large value of $T_{\phi,\Omega}$ in (3) as a manifestation of $s$ quarks in the ridge. If future LHC data can verify the correlation between $(\phi,\Omega)$ and the ridge, then they not only give support to our view that is outside hydro flow, but also pose a strong challenge to any other approach for a satisfactory explanation.

\ack{This work was supported,  in part,  by the NSFC of China under Grant No.\ 11205106. We are grateful to Prof. Chunbin Yang for helpful discussions and to Dr. Xiaoping Zhang for providing us with the preliminary values of $T_{\phi,\Omega}$ from STAR's analysis of RHIC-BES data.}

\smallskip


\medskip
\section*{References}
\begin{thebibliography}{99}
\bibitem {al1}
Abelev B {\it et al} (ALICE Collaboration) 2015 {\it Phys. Rev.} C {\bf 91} 024609 

\bibitem {al2}
Abelev B {\it et al}  (ALICE Collaboration) 2014 {\it Phys. Lett.} B {\bf 728} 216

\bibitem{bl}
Schnedermann E, Sollfrank J and Heinz U 1993 {\it Phys. Rev.} C {\bf 45} 2462

\bibitem{rhphi}
Abelev B I {\it et al} (STAR Collaboration) 2009 {\it Phys. Rev.} C {\bf 79} 064903 

\bibitem{rhomega130}
Adams J {\it et al} (STAR Collaboration) 2004 {\it Phys. Rev. Lett.}  {\bf 92} 182301

\bibitem{rhomega200}
Adams J {\it et al} (STAR Collaboration) 2007 {\it Phys. Rev. Lett.}  {\bf 98} 062301

\bibitem{bes}
Adamczyk L {\it et al} (STAR Collaboration) 2016 {\it Phys. Rev.} C {\bf 93} 021903 

\bibitem{phenix}
Adare A {\it et al} (PHENIX Collaboration) 2013 {\it Phys. Rev.} C {\bf 88} 024906

\bibitem{reco}
Hwa R C 2010 {\it Quark-Gluon Plasma 4}, ed Hwa R C and Wang Xin-Nian (Singapore: World Scientific), arXiv: 0904.2159.

\bibitem{bm}
Mohanty B 2016 {\it Strangeness in Quark Matter 2016}  (to be published in this volume)

\bibitem{jp}
Abelev B I {\it et al} (STAR Collaboration) 2009 {\it Phys. Rev.} C {\bf 80} 064912 

\bibitem{cms}
Chatrchyan S {\it et al} (CMS Collaboration) 2011 {\it J. High. Energy Phys.}  JHEP07(2011)076

\bibitem{hz}
Hwa R C and Zhu L 2012 {\it Phys. Rev.} C {\bf 86} 024901

\bibitem{hw}
Hwa R C 2013 {\it Int. J. Mod. Phys.} E {\bf 22} 1330003

\end{thebibliography}
\end{document}